# OmniBOR: A System for Automatic, Verifiable Artifact Resolution across Software Supply Chains

Bharathi Seshadri[1], Yongkui Han[1], Chris Olson[1], David Pollak[2], Vojislav Tomasevic[1,3]
[1] Cisco Systems  [2] Spice Labs  [3] Syrmia

## Abstract

Software supply chain attacks, which exploit the build process or artifacts used in the process of building a software product, are increasingly of concern. To combat these attacks, one must be able to check that every artifact that a software product depends on does not contain vulnerabilities. In this paper, we introduce OmniBOR, (Universal Bill of Receipts) a minimalistic scheme for build tools to create an artifact dependency graph which can be used to track every software artifact incorporated into a built software product. We present the architecture of OmniBOR, the underlying data representations, and two implementations that produce OmniBOR data and embed an OmniBOR Identifier into built software, including a compiler-based approach and one based on tracing the build process. We demonstrate the efficacy of this approach on benchmarks including a Linux distribution for applications such as Common Vulnerabilities and Exposures (CVE) detection and software bill of materials (SBOM) computation.

## 1 Introduction

In December 2020, a complex software supply chain attack was performed against SolarWinds, Inc., a provider of network performance monitoring tools used by many organizations across the world. A routine update to SolarWinds' Orion software was hijacked by a malicious process that injected malicious code into the build process, producing built software that contained a backdoor through which systems could be exploited [1]. In post-facto analysis, if it were possible to trace every software object to the source level that the final built binary depended on, and check their provenance, it could have been possible to prevent the attack. Having such information could also help in detecting other exploits, where dependence on a known vulnerable library (e.g. log4j) can be flagged at build time [2] [3].

What if we could know, for every software artifact – binary executable, shared object, container, etc., the complete and reproducible artifact tree including all its dependencies, and we could efficiently cross-reference that against a database of known vulnerabilities before deployment? If we had had that information, could we have remediated vulnerabilities such as Log4Shell faster? Might it even help open-source maintainers identify at-risk dependencies sooner? As software supply chains grow ever more complex, with dependencies that cross organizations, national boundaries, and multiple open-source projects, developing effective ways to perform software artifact resolution for security and other purposes has become ever more important.

Efforts to create a Software Bill of Materials (SBOM) [4] [5]aims to address this need by capturing all kinds of information about components that get built into a software artifact. The information includes attributes such as name, version, supplier, components, licensing, who to contact if there is a problem and other software metadata. However, constructing an SBOM with high fidelity is hard, especially when working with just the software and without modifying the build process. Tools such as scanners come with both false positives and false negatives, meaning that they could either miss an artifact that is used in one's software or report an artifact that is not used. For example, CVE-2023-45853 reported zlib to have a vulnerability due to MiniZip distributed with zlib through 1.3 having an integer overflow and resultant heap-based buffer overflow. However, zlib does not itself depend on MiniZip, and hence this CVE can be considered a false alarm [6] when Minizip is not installed.

In this paper, we introduce OmniBOR [7], a minimalistic scheme for build tools to create a compact Artifact Dependency Graph (ADG), tracking every software artifact incorporated into each built software product. The OmniBOR toolchain populates a unique, content-addressable reference for the Artifact Dependency Graph (ADG), called the OmniBOR Identifier (OID), and can also optionally embed that into the output artifact at build time. OmniBOR is designed to consistently build verifiable Artifact Dependency Graphs (ADGs) across languages, environments, and packaging formats, with no developer effort, involvement, or awareness. Through OmniBOR, we wish to enable automatic, verifiable artifact resolution across today's diverse software supply chains [8]. Further, this complements SBOM methods, such as Software Product Data Exchange (SPDX), Open Web Application Security Project CycloneDX, or Software Identification Tagging (SWID) [9] [10] [11]. Sec. 2 includes a more detailed discussion and comparison with SBOM methods. OmniBOR has several use cases, including CVE detection, build reproducibility, and augmenting SBOM generation. We discuss and demonstrate these applications in Sec. 4 and Sec. 5.

To summarize, the novel contributions of this paper are:
- We introduce OmniBOR, an open framework for computing ADGs and for artifact resolution.

- We provide multiple approaches to implementing OmniBOR, including compiler-based solutions (LLVM, GCC, etc.) and tracing-based solutions (bomsh, etc).
- Using the ADG, we show how an SPDX SBOM can be created. We also show how a CVE search can be performed using the ADG.

We demonstrate the effectiveness and low overhead of OmniBOR on multiple benchmarks, including OpenSSL and Linux. We show that it can be used for multiple tasks including SBOM computation and CVE detection.

The outline of the rest of the paper is as follows. We begin in Sec. 2 by putting OmniBOR in the context of related work. In Sec. 3, we present the overall architecture of OmniBOR, the nature of the data generated by OmniBOR especially ADGs. In Sec. 4, we describe how OmniBOR data is generated during the build process and embedded in the software artifacts generated by a build system. Sec. 5 presents an empirical evaluation of the OmniBOR tools we have developed. We conclude in Sec. 6.

## 2 Related Work

Tools and techniques for SBOM generation (e.g. [6][7][8]) are related to our OmniBOR effort. For every artifact, OmniBOR ADGs only record identifiers and do not include any metadata. However, SBOMs tend to record several metadata to cover provenance, origination, build environment information such as vendor, release version, license, copyright, etc. If OmniBOR Identifier (OID) is considered as one such metadata in the SBOM, it can also be seen that the SBOMs in turn provide the metadata for the OmniBOR ADGs. We argue that OmniBOR can help increase the precision of SBOMs. For example, different tools can produce different SBOMs for the same software artifact. Suppose *sbom-tool1* uses CPE for naming a component in SBOM and *sbom-tool2* uses PURL instead and *sbom-tool3* uses a custom name to identify that same artifact in the context of the product's marketing. In this case, the OmniBOR ID could be the unifying metadata in the SBOMs amidst the different naming schemes. Additionally, there is no way to represent in SBOMs today the granularity of dependencies ranging from components upto the source files. OmniBOR provides this granularity from the highest level (packages) to the lowest level (source files), especially since problems such as CVEs originate at the level of source files and hence needs to be detected at this level for higher accuracy.

Additionally, tools that operate on packaged software to produce an SBOM, such as Syft [12], do a good job of scraping the information contained in a package file and producing an SBOM with the basic data elements. However, they cannot express detailed provenance and relationship information present during the build process since that information does not persist into the generated artifact. In contrast, our results demonstrate how metadata thrown off as part of an OmniBOR-enabled build augments the SBOM generation process by capturing ephemeral elements present in the build and using those elements to create additional SBOM relationships. Our approach adds a reference in the SBOM to the root of the ADG (representing the entire unit of software) capturing build-time and run-time dependency information.

Recently, researchers have identified reproducible builds as an important component in ensuring software supply chain security, with computation of transitive dependencies seen as a key problem [13]. The NixOS project [14], which seeks to enable reproducible builds, has a tool sbomnix that generates SBOMs and dependency graphs for Nix targets; however, all the tooling is specific to Nix packages and targets and does not generalize to other build systems.

## 3 Architecture of OmniBOR

The creation of an Artifact Dependency Graph (ADG) is the basis for OmniBOR. Let us look at the building blocks that make up the ADG.

### 3.1 Artifact

An artifact is any software object of interest like a source code file (in any language), object file, shared object file, java class file, executable file, container image, etc. There are two types of artifacts. Artifacts produced by a tool are said to be *derived artifacts*, e.g., an object or executable file. Artifacts which are not derived artifacts are said to be *leaf artifacts*, e.g., source files (.c, .java, etc) constructed by hand by humans. If a source file is generated by a tool (e.g, patch), it is still considered as a derived artifact. For example, in Figure 1: Simple Example Code, foo.o is a derived artifact derived from foo.c and foo.h using *gcc*; foo.exe is derived from foo.o and bar.o using *ld*. An artifact is said to be either an input artifact or output artifact. A build tool is said to consume input artifact(s) to produce an output artifact.

```
// hdr.h
int add (int, int);
int sub (int, int);
// add.c
#include "hdr.h"
int add (int a, int b) { return (a+b); }
// sub.c
#include "hdr.h"
int sub (int a, int b) { return (a-b); }
```

*Figure 1: Simple Example Code*

OmniBOR aims to capture the dependent input artifacts during the build process and build a verifiable Artifact Dependency Graph for a software artifact. The OmniBOR standard defines three concepts, which together enable the consistent, reproducible, and embeddable encoding of the exact inputs used to build a software artifact: *Artifact Identifiers*, *Artifact Input Manifests*, and *Artifact Dependency Graphs*.

## 3.2 Artifact Identifier

Artifact Identifier or Artifact ID is a content-based identifier of a single input (for example, a single file) used to build a software artifact. Identifiers are reproducible, meaning two individuals will always derive the same identifier for the same input. With these identifiers, we can consistently and precisely identify any software artifact or its input, for use in forensics, accounting, and vulnerability management.

Artifact IDs should have the following characteristics:

- *Canonical:* Independent parties, presented with equivalent artifacts, derive the same artifact identity.
- *Unique:* Non-equivalent artifacts have distinct identities.
- *Immutable:* An identified artifact cannot be modified without also changing its identity.

Most source code artifacts are already stored using version control systems such as git and indexed by their git object identifiers (*gitoids*) as git objects of type "blob" [15]. The gitoid of an object is computed as a cryptographic hash using either the SHA1 or SHA256 algorithms. For this reason, OmniBOR has chosen to use the "gitoid" of an artifact as its Artifact Identifier. Source code leaf artifacts are thus identified via their gitoid. The same gitoid can also be used to identify vulnerable artifacts. The gitoid of an artifact can also be computed using the command *'git hash-object <artifact>'*. Git currently supports gitoids computed with both SHA1 and SHA256; when git fully standardizes to use SHA256, the support for SHA1 could be dropped from OmniBOR too [16]. We further note that OmniBOR co-exists with, but does not require, version control systems such as git.

## 3.3 Artifact Input Manifest

Artifact Input Manifest, also known as Input Manifest (IM) is a file that lists the Artifact Identifier of every input used to produce an artifact. An IM describes the immediate children of an artifact in the ADG. For example, if an executable is compiled by linking together a collection of object files, the Artifact Identifier of every object file would be listed in the Input Manifest for the executable. IMs can be identified by treating them as artifacts and applying the same identifier heuristic to them as applied to any other artifact. For a given artifact, the OmniBOR ID (OID) of that artifact is simply the gitoid of its corresponding IM. The OID is also known as Input Manifest Identifier or IMID. The OID can also be embedded directly into executable files or can be provided in a separate file alongside the artifact whose inputs they describe. A leaf artifact in an IM is represented as

*blob␣${artifact id of the child}\n*

A derived (non-leaf) artifact in an IM is represented as

*blob ${artifact id of child} bom ${OID of child's IM}\n*

However, if the derived artifact is computed using a build tool that does not generate OmniBOR data, then this artifact is represented like that of a leaf artifact (without the 'bom' field).

For example, consider the test case in Fig 1, where a shared library is built as below.

$ clang -c add.c sub.c

$ clang -fuse-ld=lld add.o sub.o -shared -o libmath.so

There are three output artifacts that are produced as a result of the above build steps – add.o, sub.o and libmath.so. For each of those output artifacts an IM for each hash type is created and placed under $OmniBOR_DIR. As mentioned in Sec 3.2, *git* currently supports both sha1 and sha256. An IM is created for every supported gitoid hash type. For example, an IM is computed each for SHA1 and SHA256 and therefore there exists an OID for SHA1 and SHA256. OID is 20 bytes for SHA1 and 32 bytes for SHA256. Supporting multiple hash types would incur increase in both computation time and storage. Therefore, it may be preferable to standardize on just one hash type, SHA256 in this case as SHA1 is weak and expected to be phased out.

```
gitoid:blob:sha1
blob 9bf37f7f0ee6005d4b8fa43f651777904dd418f1
blob e161eff37821de6b7a96f765020d182e18e46ceb
```

*Figure 2a: IM for add.o*

```
gitoid:blob:sha1
blob 7d638576ac3a7caadda3ccc31b4bc3616003a5c9
blob 9bf37f7f0ee6005d4b8fa43f651777904dd418f1
```

*Figure 2b: IM for sub.o*

```
blob 1220793697ad532c6b5b961dff012854a8b2ee24
blob 21aa6f286ac7b15cf858212c8b9573c41d754a66
blob 237dbd85835141fc04014baa6d3daba5095a8293
blob 37040d2756deb99bf195c612fe5a87cd3ce3bb64
blob 3f7f300de3ce82a28c42c9322a977186279c3d21
blob 53698e412779edc2d5ba2c533e3b5b45f9192b61
blob 679ce65c8b2ea9669c53bc398f1d888571b1bde8 bom
2b83dbc6af3153d3bfac908be5e454985ef3b6a5
blob 70a33152f9c29d07036a7abcb177c20ed40b34ed
blob 81c73b3042d00786707552a6135f6c6663608db7
blob c25c96f2744b1062c498ecb58a60d51d750c14bc
blob c8e92242fc5b32888222e279088df08e791bd52c
blob dcfc8df4b3b19f4f7541c3db5a7ff20d00c7aeda bom
d0a8c1f725250228f6048f3299f1f15797437445
blob fea351cd56632e425aa9d2a114bbf49332e39ad1
```

*Figure 3 IM for libmath.so*

For example, the IM for add.o and sub.o (see Fig 2a and 2b) contains two entries corresponding to the gitoids of the input dependencies for add.o, namely add.c and hdr.h sorted in lexigographic order. The header 'gitoid:blob:sha1" or "gitoid:blob:sha256" indicates the hash type used in the IM. Similarly, the IM for sub.o contains two blob entries with the gitoids for sub.c and hdr.h. The IM for libmath.so in Fig. 3 contains the blob entries that correspond to add.o and sub.o. In addition to these inputs, libraries such as libgcc_s.so, libc.so, crti.o, crtbeginS.o, crtendS.o, crtn.o are also used in the final linking step to produce libmath.so. Object files and static archives used during the link are all counted towards input dependencies for that artifact. However, shared objects are not counted as dependencies and are excluded from the IMs. The reason for this is that shared objects can be different at runtime as long as the libraries used during build time and runtime are API compliant. Thus, OmniBOR does not compute *dynamic* dependencies. However, shared libraries could be included as part of OmniBOR metadata though to track any dynamic dependencies if required by an application. In Fig 3, note that only the entries corresponding to add.o and sub.o have the link to their IMs specified after 'bom'. This is because the system libraries used for this experiment were not built using OmniBOR and hence the IM for those artifacts are missing. Each of these IM files are named using its OID or IMID and stored in a directory specified by the environment *OMNIBOR_DIR* or through an option to the build tool.

## 3.4 Artifact Dependency Graph (ADG)

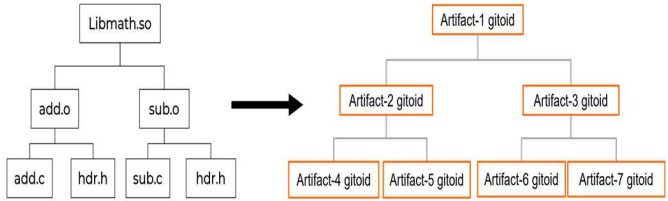

*Figure 4: Example Artifact Dependency Graph*

The Artifact Dependency Graph is a Merkle tree of hashes computed recursively starting from all the input artifacts that are transformed by a build tool into the final built artifact. It includes the direct input artifacts and the recursive set of artifacts to each input artifact all the way down to its source files. The root of the ADG is the Input Manifest Identifier of the output artifact of the build process. Each node in the ADG represents an artifact with its gitoid. The child nodes of a node/artifact in the ADG are the immediate dependencies of that artifact. Fig 4 depicts an ADG for shared library in C corresponding to the test case given in Fig 1. The root of the ADG is the shared object and it is obtained by linking two object files add.o and sub.o. These object files in turn are produced by compiling add.c and sub.c respectively with a dependency on hdr.h. libmath.so, add.o and sub.o are derived artifacts and add.c, sub.c and hdr.h are leaf artifacts. Fig 5 shows the ADG annotated with the IMs.

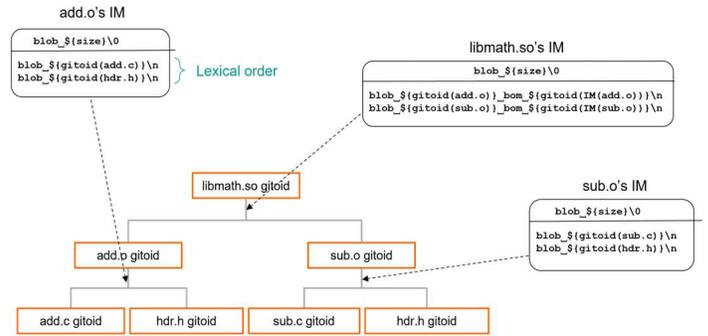

*Figure 5: ADG with Input Manifests*

Artifacts produced by different toolchains could well be different and hence the gitoids of derived artifacts are not guaranteed to be identical. For example, a binary artifact that is compiled using an optimization level O2 would not be identical to the one produced by O1 and because their gitoids will not be identical, the ADG would not be identical as well. However, the gitoids of the leaf artifacts (e.g, source files) obtained from pristine sources are expected to be identical even if the derived artifacts in the ADG are obtained by using different flavors of the build. Fig 6 shows an example where file foo$^1$.c is patched using foo.patch. Both foo$^1$.c and foo.patch are input dependencies for this patch step. The output artifact produced as a result is foo$^2$.c. By tracking every build command used in the build process, a complete ADG is thus generated. For example, an organization could check out the sources from an official open-source repository. The pristine sources would form the leaves of the ADG. Assuming the organization develops custom patches that are patched on top of the pristine sources, OmniBOR enabled 'patch' tool would generate IM for the patch step. The patched sources would now become a derived artifact and the pristine sources are the leaf artifacts. So, if a CVE is reported for the gitoid in the pristine sources, it can still be looked up from the ADG.

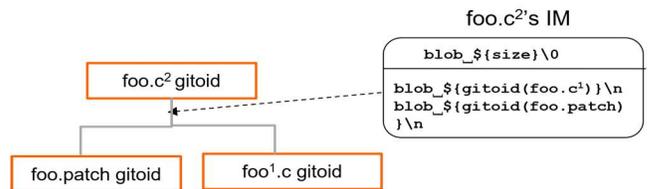

*Figure 6 ADG for patch command*

## 3.5 Embedding OmniBOR ID

In the OmniBOR system, build tools embed the OmniBOR ID (OID) into each derived artifact that is built. For ELF objects and binaries, this would mean embedding the OID in an elf section named *'.note.omnibor'* as shown in Fig 7. The .note.omnibor is an ELF section of type SHT_NOTE, with attribute SHF_ALLOC so that this section persists in the executable. This section is extensible such that it can contain OID for multiple hash types. Currently there is an entry for each for SHA1 and SHA256 hash types. An ELF Note section has 5 fields:
(1) *type*: a constant (enum) "NT_OMNIBOR" (4 bytes).
(2) *size*: specifying the size of 'Owner' string (4 bytes).
(3) *owner*: string "OMNIBOR" (8 bytes).
(4) *descriptor size*: Length of the gitoid (4 bytes).
(5) *descriptor*: Value of the gitoid (20 bytes for sha1 and 32 bytes for sha256).

This section is of size 92 bytes. A new section header for *.note.omnibor* consumes another 32 bytes. Embedding OID in the ELF object can potentially increase the size of an ELF object file by 122 (92+32) bytes. There could be +/- difference in size owing to padding adjustments for page alignment. In practice, we observed that adding *.note.omnibor* section increases the size by < 122 bytes as it fits well within the space which would otherwise be wasted due to padding adjustments.

For a java class file, this would mean embedding the OID into an annotation named @BOM in the .class file. For a generated source code file, this would mean embedding the OmniBOR identifier into the source code file using a comment. The tool *patch* would embed the OID in the source file using a comment appropriate for that file type. The embedded OmniBOR comments would then be read by OmniBOR enabled toolchain and used in the generation of IM.

OID embedded in the artifact gives us the root note of the artifact's ADG. However, any application using OmniBOR data would need to access the complete ADG and associated IMs. Embedding the entire ADG in the artifact may not be feasible due to size limitations. We note that, recently, cryptographic schemes to improve the embedding of ADG metadata in the built binary have been proposed [17].

```
$ llvm-readelf -n vmlinux
Displaying notes found in: .note.omnibor
 Owner        Data size       Description
 OMNIBOR      0x00000014      NT_GITOID_SHA1
 SHA1 GitOID: 4b6dc93e51a1b7506f40f408275e91acfd180d2c
 OMNIBOR      0x00000020      NT_GITOID_SHA256
 SHA256 GitOID: 5bc8896950cf038a3c9593b13c62257ba2e84
 e4e880591708f4e4305bf1f17cd
```

Figure 7: Sample .note.omnibor section

*Non-embedding Mode:* If the embedding of OID in the artifact is not preferred, then an alternate mechanism or protocol needs to be defined to externally establish the mapping between the output artifact and its IM. One possibility is to create a file named with the gitoid of the output artifact and store the contents of the IM or the OID in it. By doing this, post software build, the IM for an output artifact can still be derived by looking up the filename with its gitoid.

### 3.6 Metadata

The IMs contain a set of gitoids (hashes) and are not very human readable. Additionally, if any auxiliary information is needed for solving specific problems using the ADG, there needs to be a way of linking such metadata to a specific IM.

To address this need, any useful metadata is recorded in a metadata file that is written into a subdirectory of the same directory where the IMs are located. That would be ${OmniBOR_DIR}/metadata/${context}/, where *context* refers to the specific build tool that generates this metadata. Some examples of *context* are gcc, clang, go, rustc, etc. The metadata could include the names and path of the input dependencies, build command that is used to generate the artifact. Note that by using a different compiler optimization, the object file produced would be different. Two source files when compiled using different compiler optimizations produce different output artifacts. In such cases, having insight into the build commands could be very useful.

## 4 OmniBOR Data Generation and Use

The tooling for OmniBOR can be classified under two categories – one that generates OmniBOR data (*bomsh, gcc, clang, ld, lld, patch, ar, objcopy, etc*) and another that consumes this data to solve meaningful problems **[18]**. The former are termed as Producer Tools while the latter are termed Consumer Tools (see Fig. 8).

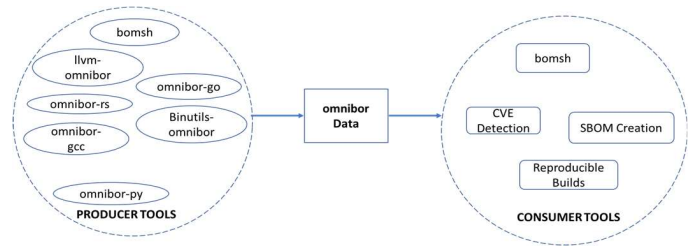

Figure 8: OMNIBOR Tools

### 4.1 OmniBOR Data Generation

The OmniBOR data can be generated in multiple ways. We describe two ways of generating this data. Both approaches are straight-forward and work with existing build systems.

### 4.1.1 Generation with Compiler (Build) Tools

The build toolchain (compilers, linkers, code generators) determines precisely what goes into executable software. OmniBOR is designed to capture this data from the build instance. We envision all tools used during a software build to become capable of generating OmniBOR data. Some examples of build tools are *gcc*, *clang*, *binutils (ld, ar, objcopy)*, *rust* compiler, *go* compiler, *pyc*, *javac*, *patch*, etc. OmniBOR data generation can easily be turned ON by setting an environment variable OmniBOR_DIR or a build tool specific option specifying the location to store the OmniBOR data. If neither option is used to specify the location to store OmniBOR data, no OmniBOR data will be generated.

For each output artifact that a build tool produces, it is expected to produce two outputs for OmniBOR: (a) Create IM and metadata files in the desired location as specified by environment variable OmniBOR_DIR or build tool option, (b) Embed the OID for this artifact within the artifact as appropriate for that artifact type. If embedding is not done, additional information needs to be logged in a file external to the artifact to derive the OIDs after the build is done. How does the build tool create the IM? Whenever a build tool opens an artifact during the build process, that artifact is considered a dependent input artifact. For example, when producing an output artifact, the compiler treats any header files it opens as an input dependency. Similarly, if a linker opens an object file or an archive file, it is considered as an input dependency for that output artifact and the gitoid of that input artifact is recorded in the IM for that output artifact. The IM file is named using its OID, the gitoid of its contents, that is `git hash-object <IM>`.

Our approach for *compiler-based OmniBOR data generation* (currently implemented for clang, lld, gcc, ld for C/C++) computes the IM as follows (also see Fig. 9) [19] [20] [21]:
1. Collect all input dependencies (.h, .c, .o, .so, linker script). For example, one can generate these with the -MD compiler option or –dependency-file linker option.
2. Compute the gitoid of the input dependencies from step 1 and sort them in lexicographic order. This set of gitoids become the contents of IM. As mentioned in Sec 3.3, there exists an IM for every supported gitoid hash type. This step records any needed metadata (like build command and path to the files) as well.
3. Compute the OID (gitoid of the IM contents from step 2) from SHA1 and SHA256 IMs. We now have 2 OIDs, one for each hash type.
4. Name the IM files. The SHA1 IM file is named as ${OmniBOR_DIR}/objects/gitoid_blob_sha1/${OmniBORId:0:2}/${OmniBORId:2:}, and the SHA256 as ${OmniBOR_DIR}/objects/gitoid_blob_sha256/${OmniBORId:0:2}/${OmniBORId:2:}.
5. Compute the gitoid of the output artifact and store the metadata file named as ${OmniBOR_DIR}/metadata/${context}/${OutputArtifactId}.
6. Embed the OID (gitoid of IM of output artifact) in the output artifact (e.g. in .note.omnibor section in ELF files).

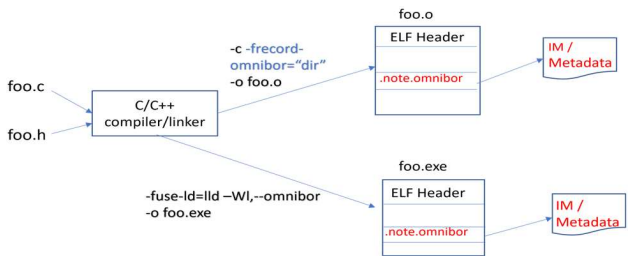

*Figure 9: llvm-omnibor Operation*

For every compilation/link command, when OmniBOR data generation is enabled, the tool produces three outputs.
- Artifact Input Manifest files
- Metadata files
- Embedding of the gitoid in the output artifact. For ELF files, it is the creation of. note.omnibor section.

In the next section, we describe an alternate approach to generating OmniBOR data using software traces.

### 4.1.2 Generation using Dynamic Tracing

We have developed a collection of tools called '*bomsh*' that does the job of OmniBOR data generation and application to various use cases [22]. To generate OmniBOR data, the tool still collects data during build but does not require specially supported toolchains. This approach relies on *strace* to determine the dependencies. The tracing script of *bomsh* is called *bomtrace*. This tool intercepts relevant build commands and generates OmniBOR data. The goal of this tool is to collect and record minimal metadata during build time and post-build completion generate the OmniBOR IMs using *bomsh* and also embed OID in the output artifacts, if desired.

*bomtrace* can trace all the processes of a software build and interpret the command lines of several commonly used build tools. The list of tools supported by *bomtrace* is listed in Table 1. The tool can be configured to interpret any or all of the tools in this list. To use this tool, a user only needs to prepend *bomtrace* to software build command as:

$ bomtrace <build-cmd>

*bomtrace* can also create IMs by computing the gitoids of the dependent input artifacts. The embedding of the OmniBOR IDs in the output artifact is also handled by this tool as appropriate. For example, for ELF objects, the embedding is accomplished using *objcopy*.

```
outfile: 29ee10adb912f7cab11c90ddc202b5f71359fc11 path: /ws/test/usenix/add.o
infile: e161eff37821de6b7a96f765020d182e18e46ceb path: /ws/test/usenix/add.c
infile: 9bf37f7f0ee6005d4b8fa43f651777904dd418f1 path: /ws/test/usenix/hdr.h
build_cmd: clang -c -o add.o add.c
==== End of raw info for PID 1016452 process

outfile: 3bc85e9f0a84e0b7bf0d1371e7c9af070c40e6d6 path: /ws/test/usenix/sub.o
infile: 7d638576ac3a7caadda3ccc31b4bc3616003a5c9 path: /ws/test/usenix/sub.c
infile: 9bf37f7f0ee6005d4b8fa43f651777904dd418f1 path: /ws/test/usenix/hdr.h
build_cmd: clang -c -o sub.o sub.c
==== End of raw info for PID 1016453 process

outfile: cfac12f47f1e1964deab731b770d7ddf0f7e0732 path: /ws/test/usenix/lib-math.so
infile: 232fd2c41d204d23899069fc89e6516aab57421b path: /lib/../lib64/crti.o
infile: 4a1908b0a92b94749c51a44252db66909b290235 path: /opt/rh/gcc-toolset-12/root/usr/lib/gcc/x86_64-redhat-linux/12/crtbeginS.o
infile: 29ee10adb912f7cab11c90ddc202b5f71359fc11 path: /ws/test/usenix/add.o
infile: 3bc85e9f0a84e0b7bf0d1371e7c9af070c40e6d6 path: /ws/test/usenix/sub.o
infile: bb5e69c89e0f4a130162afdb7376564e4afd0b4c path: /opt/rh/gcc-toolset-12/root/usr/lib/gcc/x86_64-redhat-linux/12/crtendS.o
infile: 3d5810339f0b219eb80dfa7cbd8883c3ef944351 path: /lib/../lib64/crtn.o
infile: 30976714e123aadc5fcbdca8bd9ffc25439cb4c8 path: /opt/rh/gcc-toolset-12/root/usr/lib/gcc/x86_64-redhat-linux/12/libgcc.a
dynlib: 9db7000ac1db1300a99704ff4ccf76f395e203aa path: /opt/rh/gcc-toolset-12/root/usr/lib/gcc/x86_64-redhat-linux/12/libgcc_s.so
dynlib: 4c28df352e02461d8d64482a670052a4fdc17032 path: /lib/../lib64/libc.so
dynlib: 9db7000ac1db1300a99704ff4ccf76f395e203aa path: /opt/rh/gcc-toolset-12/root/usr/lib/gcc/x86_64-redhat-linux/12/libgcc_s.so
build_cmd: clang -o libmath.so -shared add.o sub.o
==== End of raw info for PID 1016454 process
```

*Figure 10: bomtrace metadata for Example in Fig. 5*

For every command that is processed by *bomsh* during the build process, the output is a concatenated log file containing minimal and necessary metadata. The metadata contains the following types of entries:
*infile*: gitoid of input dependency
*outfile*: gitoid of output artifact
*path*: absolute path of the artifact (in ascii text)
*dynlib*: gitoid of a dependency that is a shared object
*build_cmd*: the command used to build the output artifact

For the example of Figure 1, the information recorded by *bomtrace* is given in Fig. 10. Each entry in this log contains one of the information types given above. With this information, a post processing tool can optionally generate the IMs for each output file and embed the OID in them. Doing this post builds can help keep the actual build times to a minimum.

In Linux, when a new process is created, the EXECVE syscall is invoked to load the executable into the memory and run. This EXECVE syscall can be intercepted by the *strace* tool. We analyze the created process to find out the output file and input files for this process. Based on this analysis, we establish build dependency between the output file and the input files and generate the OmniBOR IMs. Since only the EXECVE syscall is of interest, the newly introduced *strace "--seccomp-bpf"* option is a perfect choice for this purpose. This option creates a BPF (Berkeley Packet Filter) filter in the kernel so that the traced process stops only at syscalls of interest. This has significantly reduced the overhead of strace where previously all child processes were also traced.

Now we present some details on how *bomtrace* works to generate OmniBOR data. For example, when a *gcc* command is traced, *bomtrace* obtains the full command line and analyzes the command line to get a list of files to process by this gcc command. The output file is also determined from the gcc command line. *bomtrace* establishes build dependency relationship between the output file and the list of input files. To obtain the compiling dependency, the "-MD -MF depfile" option is utilized. If -M option is already being used in the *gcc* command line, the generated dependency file is read to get the dependency list. If there is no -M option usage in the *gcc* command, -MD option is added to the *gcc* command, to generate a temporary dependency file. After the dependency list is read, this temporary dependency file is deleted. Therefore, the compiling dependency is obtained for a *gcc* command. One caveat at this time is that if -MMD (an option that excludes system header files) is used, the dependency list generated by *bomtrace* uses this information only and the system headers are excluded from the IM. However, this can be mitigated if *bomtrace* runs a preprocessing command to obtain the dependency information using the -MD command. Running an additional preprocessing command alone is not likely to attribute to a larger overhead in build time.

For the linker *ld* command, all the object files and shared object files used for the link are available in the command line, so we can get the list of input artifacts directly from the command line. Note that *ld* (from 2.35v) and *lld* (from 12.0) provides a new --dependency-file option, which is similar as the "-MD -MF depfile" option for compiling. However, it seems sufficient and more efficient to analyze the *ld* command line directly to obtain the list of input files and output file.

Another example, when the ar command is run as below,
*$ ar cr libopenosc.a openosc_map.o openosc_support.o*
the *ar* process creation is intercepted by strace, and the *ar* command line is analyzed to find out the list of input files for the output archive file. In this case, the output artifact is libopenosc.a file, while the input artifacts are openosc_map.o and openosc_support.o files.

Similarly, when the strip command *"strip libopenosc.so"* is run, *bomtrace* knows that libopenosc.so is both the input and output file, and it computes the hash of libopenosc.so file before the strip command is run and the hash of the same file after the command completes, to create the IM for the strip command.

The tool can also handle some special cases. For example, consider the case in Fig 11, when the Linux kernel is compressed as a .bin.gz file and put into a piggy.o object file by including it in an assembly code and then packaged into the final bootable bzImage file. The *bomsh* tool is able to analyze the piggy.S file and determine that the vmlinux.bin.gz file is an input artifact for the output artifact piggy.o, thus creating complete dependency graph.

```
$ more arch/x86/boot/compressed/piggy.S
input_data:
.incbin "arch/x86/boot/compressed/vmlinux.bin.gz"
input_data_end:

gcc -c -o arch/x86/boot/compressed/piggy.o arch/x86/boot/compressed/piggy.S
Output file => piggy.o
Input files => piggy.S arch/x86/boot/compressed/vmlinux.bin.gz
```

*Figure 11: Example of binary embedded in assembly*

Another example is the *patch* command. For the *"cat abc.patch | patch p1"* command, *bomsh* can recognize that the inputs for this command are the original input source file before patching and the patch file, and the output file is abc.txt file. In this case, the gitoid of the file before patching is captured in the IM as a dependency (Fig 12).

```
$ cat abc.patch
--- a/abc.txt 2023-05-06 16:15:46.263661090 +0000
+++ b/abc.txt 2023-05-06 16:15:57.579587387 +0000
@@ -1 +1,2 @@
 abc
+abc
$ cat abc.patch | patch –p1
Output file => abc.txt
Input files => abc.txt abc.patch
```

*Figure 12: Example usage of patch command*

### 4.1.3 Build tools vs Tracing approaches

There are multiple ways to generate OmniBOR data for a given application. We described above two approaches – data generated by *OmniBOR-aware build tools* and a *dynamic tracing* approach using *bomtrace*, to intercept build commands and determine dependencies based on the specific build tool being invoked. The two approaches can be used in combination as well. The following are some of the key factors that influence either of the approaches.

*Ease of Use:* Both approaches are straight forward to integrate into an existing build system. For the build tools, it can be trivially integrated into an existing build process by simply adding a command line option or setting an environment variable. For the latter, the build step should be prefixed with *bomtrace*.

*OmniBOR Support:* In an ideal state, all build tools would be OmniBOR aware and generate OmniBOR data. But a downside of using build tools to generate OmniBOR data is that every version of compiler or linker or other tools used for the software build should support OmniBOR, without which there will be reduced coverage leading to gaps in the ADG. In any organization, several versions of toolchains are likely to be used across different products and releases, and the toolchains could be both open source and custom. For example, if one product uses clang version X and another product uses clang Y and another uses *gcc* version Z, all of X, Y and Z tools should be capable of supporting OmniBOR. This can pose a practical challenge to make OmniBOR support available for all the toolchains used to build an application. In cases, where older toolchains are being used and toolchain upgrades are not practical, this would mean losing OmniBOR data from such builds. However, the same *bomtrace* can support several build tools under a single framework. For example, *bomtrace* is the same whether it needs to support *clang-X, clang-Y* or *gcc-Z* compilers. This can serve as a huge incentive for any software builds to begin adopting OmniBOR where build tool support is not possible. Table 0 lists the build tools for which some form of OmniBOR support is available as of now.

| Tool | bomsh | Build-tool |
|---|---|---|
| Gcc/ld, clang/lld, gnu patch | Y | Y |
| Rust, go | Y | Y* |
| Java, Binutils(Ar, objcopy, strip) | Y | X |
| eu-strip, chrpath, ranlib, objtool, debugedit, sortextable, sorttable, resolve_btfids, dwz, dpkg-deb, rpmbuild, sign-file, sepdebugcrcfix | Y | X |

*Table 0 List of OmniBOR supported build tools (Y means support available, X means not available) *initial prototype available*

*Generated OmniBOR data:* The ADG created from OmniBOR data generated by an OmniBOR aware build tool or *bomsh* tools are expected to be very similar as both approaches are build-based and infer data from the actual build. However, there could be some differences in the contents of IMs arising from system header files and/or certain shared objects. With additional testing, the implementation could be adjusted to bridge any gaps.

*Runtime Overheads:* The OmniBOR build tools record the dependencies and generate IMs during the build process itself. This introduces lesser overheads when compared to bomsh that does an additional work of tracing dynamically the entire build process and determining the dependencies external to the build tool. See Sec 5.1 for comparison of build time overheads with both the approaches.

*Maintenance of the OmniBOR support:* If a new OmniBOR feature is needed, such as recording a new metadata type or an introduction of a new hash type, the change needs to be propagated in all build tools that support OmniBOR. On the other hand, *bomsh* would be updated once but can be used by all the build tools supported by *bomsh*.

## 4.2 OmniBOR Consumer Tooling

In this section, we will describe the tooling developed for CVE Detection and Reproducible builds. Irrespective of how the OmniBOR data is generated, if it is as per the OmniBOR spec [23], a common set of consumer tooling could be leveraged.

### 4.2.1 CVE Detection

The OmniBOR data generated by build tools can greatly help with software vulnerability detection.

One approach is to build a vulnerability database indexed by gitoids. All the gitoids of the known source files with a vulnerability (resolved or otherwise) are entered into this database and tagged with metadata or attributes of CVEs. Once the ADG is built, each gitoid in the ADG is looked up in this vulnerability database. If present, the result of the lookup reports all CVEs that apply to the built software.

We have developed some tools to automate this procedure. For example, a *bomsh* script can be used to create such a vulnerability database from the source repository, e.g., the git repository of the Openssl library. From the metadata collected during an OmniBOR enabled build, another bomsh script can be used to create the OmniBOR ADG, and then bomsh will utilize such a vulnerability database and the OmniBOR data to report vulnerabilities for artifacts of the software [24]. It not only tells you if the software is vulnerable, but also points out which source file artifact is the root cause of the vulnerability. In addition, more details like the full build command chain are provided to know how this vulnerable source file is built into the software. If the software has already fixed the CVEs, the tooling also reports the CVEs that are fixed.

Another approach is to utilize the SBOM documents generated by bomsh (See 5.2). These SBOM documents include package name, version, pURL, which can be fed to a CVE database to search for vulnerabilities. Existing CVE databases mostly associate CVEs with known package name/versions, thus this approach can work well with most of the existing CVE databases.

The first approach is more accurate than the second approach as it can track vulnerabilities at the source file level. For now, we lack a production-grade vulnerability database, thus its usage is still limited. However, there is a growing community that has started working on such a new vulnerability database, such as https://osv.dev (See also 5.3.2). The second approach is less granular than the first approach, but it covers more use case scenarios because the existing CVE databases are built/organized based on package name/version. Therefore, both approaches are very useful for detecting vulnerabilities in a software.

### 4.2.2 Reproducible builds

Another useful application of bomsh is with reproducible builds [25]. A reproducible build ensures that given the same source code, build environment, and build instructions, any build of the software will produce the exact same binary, byte for byte. As of today, more than 90% of Debian packages are build-reproducible. With *bomsh*, a Debian package can be reproduced byte-to-byte identical with the official Debian package, and its OmniBOR data is generated during the reproducible build. Therefore, each user can separately re-generate and verify the same OmniBOR ADG of Debian packages with *bomsh* without requiring special OmniBOR enabled tools.

*bomsh* automates reproducible build for a Debian package as follows: Given a Debian buildinfo file of an official Debian package, the script creates a *docker* container to do all the reproducible builds. Inside this *docker* container, a *chroot* build environment is created, and the exact same package versions as specified in the buildinfo file are installed in this chroot environment. The script also uses *bomsh* inside this docker container, and runs the Debian package build with *bomtrace*, thus generating OmniBOR IMs during the build. It also verifies the newly built Debian packages have the identical checksums as specified in the buildinfo file. When the script completes, the OmniBOR documents of the official Debian packages are created. All users should be able to get the same OmniBOR documents to verify it. This implies that for already-released official Debian packages, OmniBOR documents can be generated by any user later on-demand, not necessarily created during the build of the official Debian packages. This is a big advantage of reproducible builds, which can motivate the adoption of reproducible builds by other Linux distros as well. Such an application to reproducible builds becomes extremely challenging with special OmniBOR-aware toolchains. The already-released Debian packages are built with older versions of gcc/clang compilers, which do not have OmniBOR support yet. Therefore, *bomsh* can work perfectly well with already released software packages, not just newly created software.

## 5 Experimental Evaluation

### 5.1 Scalability and Overheads

We have built OpenSSL and Linux kernel with our tools and generated OmniBOR ADG using both approaches mentioned in Sections 4.1.1 and 4.1.2. All builds were performed on a Ubuntu 20.04.1 machine with -j8 parallelism.

Build tools such as gcc and clang generate metadata, IMs and embed the OID in the output artifact, all during build time. However, in the case of *bomtrace*, only the recording of metadata is done during the actual build and *bomsh* is used post build to generate the IMs and embed the OID within the output executable and shared objects.

As seen in Table 1, for an openssl build, which took 30.7s in real time with clang, there was an increase of build time by 4.2s (13.6%). A linux baseline build that took 2523s (42m3s) with clang increased by 169s (less than 3m) when OmniBOR was enabled. Similarly, the overheads observed for openssl and linux kernel builds with gcc and ld tools are 27% and 10.5% respectively (see Table 1, 2). Thus, we see that *the overheads with build-tool are fairly low*. With *bomtrace*, the openssl build time increased by 22s (over the Clang baseline, see Table 3). It also took an additional 25s to create the IMs and embed the OID in the final executables. Since it is a fast build (only ~30s), the build time overhead appears to be high at 71%. However, for a linux build, with *bomtrace*, the overhead observed was only 8.4%, an increase of 3m33s. Table 3 shows overheads seen with *bomsh*. The *bomtrace* tool only records the metadata, thereby decreasing the build time overhead to create the output artifacts. For a linux kernel build, the overhead of *bomsh* is comparable to that of a OmniBOR enabled *clang* build. The *real overheads* of *bomsh* lie in processing the metadata to produce the IMs and especially *in embedding the gitoid in the output artifacts*. We also experimented with clang generating only the metadata during build time and used the *bomsh* to generate the IMs post build using the metadata. But this did not significantly decrease the overheads observed during build time. It must be noted that typically only special production builds are expected to have the OmniBOR feature enabled and hence this overhead does not apply to normal development builds.

| Benchmark | Baseline Clang | Clang/lld for OmniBOR | Baseline gcc | Gcc/ld for OmniBOR |
|---|---|---|---|---|
| Openssl | 30.7s | 34.90s (13.6%) | 30s | 38s (27%) |
| Linux Kernel | 2523s | 2692s (6.7%) | 1777s | 1965s (10.5%) |

*Table 1: Timing Overhead: OmniBOR data generation with gcc, clang*

| Benchmark | Baseline Clang | Clang for OmniBOR | bomtrace for OmniBOR |
|---|---|---|---|
| Openssl | 30.7s | 34.90s (13.6%) | 77.6s (2.5x) |
| Linux Kernel | 42m | 45m (6.7%) | 105m (2.5x) |

*Table 2 Timing overhead for OmniBOR data generation (build + metadata + IM + ELF exe/.so embedding)*

| | Bbtrace Overhead (clang/lld as Baseline) | | | |
|---|---|---|---|---|
| | Build time/ Metadata | IM* | Embed +IM* | |
| Openssl | 52.6s (71%) | 7s (88%) | 25s[1] (1.5x) | 50s[2] (2.5x) |
| Linux kernel | 45m35s (8.4%) | 23m (63%) | 60m[1] (2.5x) | 120m[2] (4x) |

*Table 3 bomsh Overheads*    * Post build steps;   [1] – Embed ELF executables & shared obj only;  [2] – Embed all objects

The size overheads reported in Table 4 are expected to be similar for any of the OmniBOR approaches used. The size of the IMs generated for a linux build is 3GB. If we store only sha256 IMs, then the storage space can be reduced by 43%. Since the IMs are plain ascii files, they compress well. Also, note that there is negligible increase in the size of the executables and shared objects after embedding the OID. *bomsh* generates minimal raw data (1.5G for Linux) during build time and uses this information to generate IMs post build.

| Parameter | Openssl | Linux kernel |
|---|---|---|
| Size of IM (sha1 + sha256) | 32MB (14M+18M) | 3GB (1.3+1.7) |
| Compressed | 4.1MB | 1.2GB |
| #IM (sha1) | 1497 | 30186 |
| exe/ .so size (+embed oid) | libcrypto:1.4%(84kb) libssl:0.7% (8kb) | Vmlinux: No change |
| Metadata | 24MB | 1.3GB |
| Size of bomtrace Raw logfile | 19MB | 1.5GB |

*Table 4 Size overheads*

## 5.2 SBOM generation

SBOM documents endeavor to provide a standard format for exchanging data about a unit of software. What constitutes a unit of software and what data elements to include in the document are flexible to accommodate diversity in software. Installable packages (RPM, Debian, etc.) would be an example of a unit of software. Archives of source files and container images also qualify as software units. We show how bomsh can be used to rebuild Alma Linux binary packages and generate SPDX documents for them.

Basic data elements in an SBOM document generally fall into the following groupings:

- Document characteristics, including (i) who prepared the document and when; (ii) document identifier / Unique name of document, and (iii) Document subject matter.
- Software Unit Characteristics, including (i) Name / Version / Other identifiers; (ii) Items that constitute the software unit (Files), etc.
- Provenance, including where / how was the software acquired; what changes or modifications were made to original or dependent software, etc.
- Relationship Information, i.e., the thread binding the various data elements.

The NTIA in its "The Minimum Elements for a Software Bill of Materials (SBOM)" report [26] focused on these basic groupings.

Competing SBOM standards introduce additional elements designed to facilitate various use cases [27] [28]. Items in this area include

- Licensing
- Security / Vulnerability data
- Build characteristics
- AI (e.g. training set and model)

We selected SPDX 2.3 as our SBOM document specification. This choice was based primarily upon familiarity with the specification. For our build process, we rebuilt Alma Linux binary packages from their associated source packages and created the SBOM based on that binary package as the described unit of software. We use the rebuild of the sysstat-11.7.3-9.el8.x86_64.rpm package built using bomsh (Sec 4.2.2) in the examples below. Adding a reference to the IM for the entire unit of software was a design tradeoff. The IM contains data identifying every input file (including transitive dependencies) used to produce a file present in the package. The IM uses a very concise method for representing this data. An alternative would have been to pull all the IM data into the SBOM document. The SPDX 2.3 spec contains the necessary structures that would have allowed us to document each of those input files (including file hashes) and represent the relationship of those files in the SBOM. This, however, would have led to a much larger and more complex SBOM. It wasn't clear how often someone would need to have all the file information immediately available (by putting it in the SBOM) as opposed to referencing the material only when package name and version indicated a potential problem. We went with the later (layered) approach whereby initial vulnerability decisions would be based on package name and versions. Detailed analysis could then be performed with the referenced IM.

Fig 13 shows the IM reference included as an "externalRefs" in the "packages" section of the SPDX 2.3 SBOM. The hash value in the "referenceLocator" is the "*git hash-object*" of the IM of the package RPM. The IM for the package was archived using that hash as the lookup key. A reference to the pURL for that package helps associate the gitoid with pURL. Using pURL, we can also look up vulnerabilities in the package from public databases such as [29].

```
"packages": [
  {
    "SPDXID": "SPDXRef-Package-sysstat-ca62b83c028692e2",
    ...
    "externalRefs": [
      {
        "referenceCategory": "PERSISTENT_ID",
        "referenceLocator": "gitoid:blob:sha1:
31806346cd517fe1b94ed936206136575fbb3c5e",
        "referenceType": "gitoid"
      },
      {
"referenceCategory": "PACKAGE_MANAGER",
"referenceLocator": "pkg:rpm/almalinux/sysstat@11.7.3-
7.el8?arch=x86_64&distro=almalinux-8.9&upstream=sysstat-
11.7.3-7.el8.src.rpm",
"referenceType": "purl"
      }
```

*Figure 13 IM Reference Locator as SBOM External Reference*

Software package specifications typically include run-time dependencies in their set of metadata. Thus, it is straightforward for SBOM generation tools operating on a built package to capture these relationships and add them to the final document. However, build-time dependencies aren't needed for run-time installation and therefore generally not captured.

As part of building the binary package from source with OmniBOR-enabled build tooling, we were able to capture the names and versions of all the other packages pulled in as part of the build process. This includes the packages containing dynamically linked libraries that would show up as run-time dependencies (see, Fig 14). It also includes packages containing included source files (see, Fig 15). The names and identifying information of those source files used to build the binary package would be lost were they not gathered in the referenced IM. Similarly, the information on the packages that supplied that source material would also not be present in a resultant binary package as an installation dependency. Including that build-time information in an SBOM reinforces the layered approach to software transparency. It allows one to filter on problematic upstream packages quickly and efficiently and then use the IM to drill down for specific information.

```
"relationships": [
  {...
    "relatedSpdxElement": "SPDXRef-Package-gcc-
9262b76aa882f04c",
    "relationshipType": "DEPENDS_ON",
    "spdxElementId": "SPDXRef-Package-sysstat-
ca62b83c028692e2"
  },
  {
    "relatedSpdxElement": "SPDXRef-Package-glibc-devel-
68bfa46134cb6792",
    "relationshipType": "DEPENDS_ON",
    "spdxElementId": "SPDXRef-Package-sysstat-
ca62b83c028692e2"
  },
  {
    "relatedSpdxElement": "SPDXRef-Package-lm-sensors-devel-
ce0dac65cff5fb01",
    "relationshipType": "DEPENDS_ON",
    "spdxElementId": "SPDXRef-Package-sysstat-ca62b83c028692e2"
  } ],
```

*Figure 14 SBOM Run Dependency Relationship*

## 5.3 CVE Detection

```
"relationships": [
  ...
  {
    "relatedSpdxElement": "SPDXRef-Package-gcc-
9262b76aa882f04c",
    "relationshipType": "BUILD_DEPENDENCY_OF",
    "spdxElementId": "SPDXRef-Package-sysstat-
ca62b83c028692e2"
  },
  {
    "relatedSpdxElement": "SPDXRef-Package-glibc-devel-
68bfa46134cb6792",
    "relationshipType": "BUILD_DEPENDENCY_OF",
    "spdxElementId": "SPDXRef-Package-sysstat-
ca62b83c028692e2"
  },
  {
    "relatedSpdxElement": "SPDXRef-Package-glibc-headers-
8e6880821157632b",
    "relationshipType": "BUILD_DEPENDENCY_OF",
    "spdxElementId": "SPDXRef-Package-sysstat-
ca62b83c028692e2"
  },
  {
    "relatedSpdxElement": "SPDXRef-Package-kernel-headers-
cffe7dfe565bdf4d",
    "relationshipType": "BUILD_DEPENDENCY_OF",
    "spdxElementId": "SPDXRef-Package-sysstat-
ca62b83c028692e2"
  }, ]
```

*Figure 15 SBOM Build Dependency Relationships*

OmniBOR ADG contains gitoids of all the dependencies involved in producing an artifact. If the gitoid of a vulnerable component or source file is known, this only involves a look up operation on the ADG to check if the ADG contains the vulnerable gitoid. For example, if a CVE database (DB) is available for a product, the ADG obtained from a release build can be checked for the presence of any vulnerable gitoids and this can be used as one of the gating factors before releasing the product. We demonstrate CVE detection under two scenarios: rebuilding sysstat rpm packages and checking for CVEs against a private sample CVE DB and another one where we populated a central DB of OmniBOR Corpus and demonstrate querying for CVEs.

### 5.3.1 CVE Detection in sysstat package

To prototype the idea, we constructed manually a sample CVE DB containing vulnerable gitoids tracking CVE-2022-39377 and CVE-2019-16167 using [24]. The CVE DB contains gitoids for source files that contain the vulnerability and gitoids for files or patch files that fixed the CVE as well. Fig 16 shows one instance of gitoid containing a CVE and the gitoid that fixes the CVE. We rebuilt several releases of systat packages (versions 11.7.3-4 to 11.7.3-9) from almalinux [30] using the *bomsh* mentioned in Sec 4.2.2 and constructed the ADG.

```
"ea659e471014095e4317cc658a73e40359d00562": {
    "CVElist": [
        "CVE-2019-16167"
    ],
    "file_path": "sa_common.c"
},
"ad739fd04cd1a8bcaf6564c109e2d6ec46d0a754": {
    "FixedCVElist": [
        "CVE-2019-16167"
    ],
    "file_path": "CVE-2019-16167_memory-corruption-due-
to-an-integer-overflow.patch"
},
```

*Figure 16 Sample entries in CVE DB for sysstat package*

By checking for the gitoids in the CVE DB in the ADG we could determine the list of open and fixed CVEs (see Fig 17) in each of these versions. For example, both the CVEs are found to be fixed in 11.7.3-9 version and there are no open

```
$ rebuild_rpm.py -c alma+epel-8-x86_64 --docker_image_base
almalinux:8 -s sysstat-11.7.3-5.el8.src.rpm -d scripts/sam-
ple_sysstat_cvedb.json -o outdir
  "rpms/sysstat-11.7.3-5.el8.x86_64.rpm": {
      "CVElist": [
          "CVE-2022-39377",
          "CVE-2019-16167"
      ],
      "FixedCVElist": [
          "CVE-2019-16167"
      ],
  },
------------// for 11.7.3-p9.el8.src.rpm // -------------
  "rpms/sysstat-11.7.3-9.el8.src.rpm": {
      "CVElist": [],
      "FixedCVElist": [
          "CVE-2022-39377",
          "CVE-2019-16167"
      ],
```

*Figure 17 CVE Search results for sysstat package*

```
1  "blob f9c87607a0b4e0dcc7e3f61d6d7cb557cb56a74d bom
2  bea3dfc2f451cb423672924a0a3c6ff81ffc4b9c": {
3      "CVElist": [
4          "CVE-2022-39377"
5      ],
6      "FixedCVElist": [
7          "CVE-2019-16167"
8      ],
9  "blob ad739fd04cd1a8bcaf6564c109e2d6ec46d0a754": {
10     "FixedCVElist": [
11         "CVE-2019-16167"
12     ],
13     "file_path": "CVE-2019-16167_memory-corruption-due-to-
an-integer-overflow.patch",
14     "prov_pkg": "sysstat-11.7.3-5.el8.src.rpm"
15     },
16
17 "blob ea659e471014095e4317cc658a73e40359d00562": {
18     "CVElist": [
19         "CVE-2019-16167"
20     ],
21     "file_path": "sa_common.c",
22     "prov_pkg": "sysstat-11.7.3-5.el8.src.rpm"
23 },
24 "build_cmd": "/usr/bin/patch --no-backup-if-mismatch -p0 --
fuzz=0 < /builddir/build/SOURCES/CVE-2019-16167_memory-corrup-
tion-due-to-an-integer-overflow.patch",
25 "file_path": "sa_common.c",
26 "prov_pkg": "sysstat-debugsource-11.7.3-5.el8.x86_64.rpm"
27 },
```

*Figure 18 Detailed Analysis of CVE Search results for sysstat package*

CVEs for this version. Both the CVEs were found in 11.7.3-5 version, but one of the CVEs is also fixed in this same version. By referencing the gitoids in the ADG, we can also understand how the CVE fixes came in. Fig 18 shows that for CVE-2019-16167, for original rpm package (line 22) contained the vulnerable sa_common.c (line 21) but it can be seen (line 13) that a patch file from the same package has been used to fix the issue during build. In the sysstat-debugsource package, the source file sa_common.c was derived by applying a patch (line 24) and the patched version was used for the builds. Note that the gitoid of original and patch file is recorded in the ADG when patch command (line 24) is run (See also Fig 6). Therefore, this release is also said to fix the CVE. Even though CVE-2022-39377 was reported and fixed in Nov 2022 in 11.7.3-8, we were easily able to check for the existence of this CVE in the older releases 11.7.3-5 (Nov '20) to 11.7.3-7(Dec '21) using the ADG built for those releases.

### 5.3.2 CVE Detection using OmniBOR Corpus

We built the Goat Rodeo OmniBOR Corpus (OC) that provides a {gitoid, Package URL (purl)} to artifact mapping and a bi-directional index of artifacts to related artifact [31]. Thus, with a purl, one can see all the gitoids associated with that purl (Fig 19).

```
curl https://goatrodeo.org/omnibor/purl/pkg:ma-
ven/org.apache.logging.log4j/log4j-core@2.7
{
  "identifier": "pkg:maven/org.apache.logging.log4j/log4j-
core@2.7",
  "contains": [
    "gi-
toid:blob:sha256:aecb3872c526d7c5805e73c63afee2bafed64d1f35719
fb83b2d11f6dd1cc958",
    "gi-
toid:blob:sha256:f14a09c612371efe86ff8068e9bf98440c0d59f80e09d
f1753303fe6b25dd994"
  ],
  "containedBy": [],
  "metadata": {
    "filename": [
      "pkg:maven/org.apache.logging.log4j/log4j-core@2.7"
    ],
    "purl": ["pkg:maven/org.apache.logging.log4j/log4j-
core@2.7"],
    "filetype": ["purl"],
    "other": [],
    "_version": 1
  },
  "_timestamp": 1707249662455,
  "_version": 1,
  "_type": "purl"
}
```

*Figure 19 – purl to GitOID mapping*

With the gitoid information, it's possible to look up a particular package (Fig 20) And the package contents can be queried (Fig 21).

The individual class file has a back-reference to the packages that contain the gitoid - in this case a back-reference to log4j-core 2.7. Using the back-reference indexing, the OC can be used to determine the contents of an artifact, for example a JVM JAR file, even if the artifact itself was never indexed. This allows for composition analysis of a random artifact (Fig 22).

```
java -jar goatrodeo.jar -a commons-logging-1.2.jar

commons-logging-1.2.jar, 100 %, No CVEs

java -jar goatrodeo.jar -a l4jgood.jar

log4j-api-2.22.1.jar, 98 %, No CVEs
bcel-6.8.1.jar, 99 %, No CVEs
commons-lang3-3.14.0.jar, 99 %, No CVEs
log4j-core-2.22.1.jar, 99 %, No CVEs
```

*Figure 22 – Composition Analysis*

In the above, two different JAR files are analyzed. The first, not surprisingly, has a 100% overlap with commons-logging-1.2 as the file analyzed is that JAR file. The second analysis is of an "UberJAR" or "Assembly" of custom code and the dependent artifacts. This is a common packaging mechanism for JVM code. The analysis shows that the "l4jgood.jar" file also has a 98% or 99% overlap with log4j, bcel, commons, log4j-core. There are no currently known vulnerabilities in the contained packages. Let's try the same analysis with a file that contains a vulnerable version of log4j (Fig 23).

```
curl https://goatrodeo.org/omnibor/gi-
toid:blob:sha256:f14a09c612371efe86ff8068e9bf98440c0d59f80e09d
f1753303fe6b25dd994
{
  "identifier": "gi-
toid:blob:sha256:f14a09c612371efe86ff8068e9bf98440c0d59f80e09d
f1753303fe6b25dd994",
  "contains": [
    "gi-
toid:blob:sha256:008e581aa96f69b85b9e3e9546bf699625a7eb1d4419c
2c7843a2a06a21b1b5e",
    ...
    "gi-
toid:blob:sha256:ff8f20c6fd80e7f0a0bdac5914f5b1108d612723db76d
a10e6145a7b742934b2",
    "gi-
toid:blob:sha256:ffeab6e32d803f241dc9c9bdc4d305dcfa90209ba7819
7ca8c12e79494de9071"
  ],
  "containedBy": [ ],
  "metadata": {
    "filename": [
      "log4j-core-2.7.jar"
    ],
    "purl": [
      "pkg:maven/org.apache.logging.log4j/log4j-core@2.7"
    ],
    "vulnerabilities": [
      {
        "vulns": [
          {
            "id": "GHSA-7rjr-3q55-vv33",
            "summary": "Incomplete fix for Apache Log4j vul-
nerability",
            "details": "# Impact\n\nThe fix to address [CVE-
2021-44228](https://nvd.nist.gov/vuln/detail/CVE-2021-44228)
in Apache Log4j 2.15.0 ……",
        ],
        "filetype": [
          "package"
        ],
        "filesubtype": [
          "jar"
        ],
        "other": [
```

*Figure 20 GitOID to package mapping*

```
curl https://goatrodeo.org/omnibor/gi-
toid:blob:sha256:008e581aa96f69b85b9e3e9546bf699625a7eb1d4419c
2c7843a2a06a21b1b5e
{
  "identifier": "gi-
toid:blob:sha256:008e581aa96f69b85b9e3e9546bf699625a7eb1d4419c
2c7843a2a06a21b1b5e",
  "contains": [
  ],
  "containedBy": [
    "gi-
toid:blob:sha256:f14a09c612371efe86ff8068e9bf98440c0d59f80e09d
f1753303fe6b25dd994"
  ],
  "metadata": {
    "filename": [
      "org/apache/logging/log4j/core/config/plugins/proces-
sor/PluginProcessor.class"
    ],
    "vulnerabilities": [

    ],
    "filetype": [
      "class"
    ],
    "other": [

    ],
```

*Figure 21 GitOID to build artifact mapping*

```
java -jar goatrodeo.jar -a l4jbad.jar
log4j-core-2.7.jar, 99 %, {
  "vulns": [
    {
      "id": "GHSA-7rjr-3q55-vv33",
      ...
    }
  ]
}
log4j-api-2.7.jar, 98 %, No CVEs
bcel-6.8.1.jar, 99 %, No CVEs
commons-lang3-3.14.0.jar, 99 %, No CVEs
```

*Figure 23 – Composition Analysis with CVE*

In this case, the vulnerable version of log4j is identified in the JAR file and the vulnerability information is displayed.

Using the OmniBOR Corpus, it's possible to determine the composition of software artifacts including JVM JAR files, Docker image layers, and other "things that contain open source things." With the composition analysis, the likely CVEs and other data about the artifact can be determined.

# 6 Conclusion

In summary, we have presented OmniBOR, a minimalistic scheme for build tools to create an artifact dependency graph for a built software product. We present the architecture of OmniBOR, the underlying data representations, and two implementations on top of build processes that produce OmniBOR data. The efficacy of this approach is demonstrated on benchmarks including a Linux distribution for applications such as CVE detection and SBOM computation. For future work, we plan to standardize the OmniBOR specification and also add support for OmniBOR build tools for dynamically loaded languages such as Java and interpreted programming languages such as Python, Perl and Unix shell scripts. We also believe the ADGs computed using our OmniBOR tools form a useful type of compiler-generated provenance information that can be used in a variety of security tasks such as bug detection and forensic analysis, as demonstrated by recent work [32].

# Acknowledgements

We thank the OmniBOR community and our colleagues at Cisco Systems for their inputs and support throughout this project.